\documentclass[aps,showpacs,prx,twocolumn,superscriptaddress]{revtex4}
\usepackage{epsfig,amsmath}
\usepackage{subfigure}
\usepackage{graphicx}
\usepackage{dcolumn}
\usepackage{stmaryrd}
\usepackage{mathrsfs}
\usepackage{pifont}
\usepackage{amsthm}
\usepackage{amssymb}
\usepackage{bm}
\usepackage{latexsym}
\usepackage{hyperref}

\newcommand{\beq}{\begin{equation}}
\newcommand{\eeq}{\end{equation}}
\newcommand{\beqa}{\begin{eqnarray}}
\newcommand{\eeqa}{\end{eqnarray}}

\usepackage{color}

\begin{document}

\title{Fault-tolerant breathing pattern in optical lattices as a dynamical quantum memory}
\author{Zhao-Ming Wang}
\affiliation{Department of Physics, Ocean University of China, Qingdao 266100, China}
\affiliation{Department of Theoretical Physics and History of Science, University of the
Basque Country UPV/EHU, 48008, Spain}
\author{Lian-Ao Wu \footnote{Corresponding author: lianao.wu@ehu.es}}
\affiliation{Department of Theoretical Physics and History of Science, University of the
Basque Country UPV/EHU, 48008, Spain}
\affiliation{IKERBASQUE, Basque Foundation for Science, 48011 Bilbao, Spain}
\author{Michele Modugno}
\affiliation{Department of Theoretical Physics and History of Science, University of the
Basque Country UPV/EHU, 48008, Spain}
\affiliation{IKERBASQUE, Basque Foundation for Science, 48011 Bilbao, Spain}
\author{Mark S. Byrd}
\affiliation{Department of Physics, Southern Illinois University, Carbondale,
Illinois 62901-4401, USA}
\author{Ting Yu}
\affiliation{Center for Controlled Quantum Systems and Department of Physics and
Engineering Physics, Stevens Institute of Technology, Hoboken, New Jersey
07030, USA}
\author{J. Q. You}
\affiliation{Beijing Computational Science Research Center, Beijing 100084, China}
\date{\today}

\begin{abstract}
Proposals for quantum information processing often require the development of new quantum technologies.
However, here we build quantum memory by ultracold atoms in one-dimensional optical lattices with existing state-of-the-art technology. Under
a parabolic external field, we demonstrate that an arbitrary initial state at an end of the optical lattices can time-evolve and revive, with very
high fidelity, at predictable discrete time intervals. Physically, the parabolic field, can catalyze a breathing pattern. The initial state is ``memorized" by the pattern and can be retrieved at any of the revival time moments.  In comparison
with usual time-independent memory, we call this a {\em dynamical} memory.  Furthermore, we show that the high fidelity of the quantum state at
revival time moments is fault-tolerant against the fabrication defects and even time-dependent noise.
\end{abstract}

\maketitle


\section{introduction}
Quantum information requires practical setups in order to be utilized. The setups,
from a simple quantum memory to a universal quantum computer, can be theoretically abstract, but are expected to be implementable
using physical systems. For these idealized designs it is often difficult to find practical realizations with state-of-the-art technologies (e.g., the design of perfect state transfer, PST\cite{PST}).
Thus we often yearn for new or different technologies. On the other hand, even though quantum technologies have been developed rapidly in recent years, it is often unclear if these technologies are compatible with the theoretically idealized designs.
It appears that {\em out-of-the-box ideas} which are compatible with existent technologies are desired. We therefore ask ourselves: can a physical entity, controlled with {\em existing} technologies perform the {\em same functions} as these idealizations given that we are allowed to look
from different perspectives? That is, we wish to to find practical physical processes (from existing quantum technologies) that reproduce, partly or completely, what an idealization is able to do.  Here we will put this idea into practice for a simple design--quantum storage--by describing a quantum memory that uses existing optical-lattice technologies.
The term dynamical quantum memory, compared with the conventional quantum memory, means that the state will time-evolve but revive after certain evolution times. The current implementation with ultracold atoms in optical lattices has the advantage that these system allow a fine tuning of the relevant parameters (as the site couplings or the external magnetic field) \cite{bloch} and a precise control at the single site level \cite{wuertz2009,bakr2010,Weitenberg}.

Any quantum apparatus requires reliable physical entities serving as long-lived quantum memories in noisy environments.
General strategies for protecting a quantum state include the quantum
error correction codes \cite{shor1995,bloch2009,ekert1996,gottesman1996} and dynamical
coupling pulse control
\cite{Knill2000,Lidar1998,Viola2,Uhrig1,Uhrig2,Lidar,Wu09}. Theoretical proposals are numerous, and use
for instance, photon states \cite{Lukin,Fleischhauer}, a state in free nuclear ensemble \cite{Taylor, Poggio}
. Recently, quantum storage through using non-perturbative
dynamical decoupling control based on the quantum state diffusion approach \cite{Jing2012} and the spin chain
model \cite{Wang2012} have been investigated.  Experimental demonstrations exist for simple systems such as light or atoms \cite{light,EXP}.

Using existent technology in one-dimensional optical lattices, we demonstrate that an arbitrary
quantum state initially at one end can revive itself at predictable discrete intervals with very high fidelity. The
breathing pattern occurs in the intensive parabolic external field regimes and is quasi-periodic. Remarkably,  the pattern is fault-tolerant
against fabrication defects and even time-dependent noise. A quantum state can be stored in the dynamical process at these discrete time moments, in other words,
the pattern carries a dynamical memory. Interestingly, the setup becomes that of quantum state transfer along the $XY$ chain \cite{Liu}  when we turn off the external field.

\begin{figure}[t]
\centering
\includegraphics[width=0.9\columnwidth]{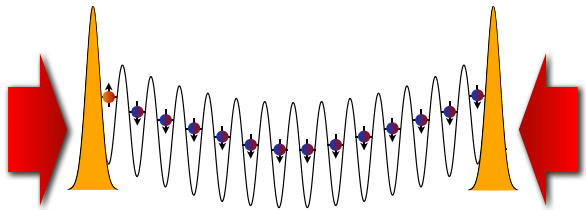}
\caption{(Color on line) Sketch of the experimental setup. Two counter-propagating laser beams (red arrows) create the 1D optical lattice; two additional transverse laser beams
(yellow gaussians) may be used to open the chain. The parabolic modulation of the on-site
energies can be realized by an additional harmonic confinement or even by the same laser beam producing the lattice. The whole setup could even be created by means of a spatial light modulator (SLM) technology, that in principle allows one to design arbitrary potentials for ultracold atoms (see \cite{Wang20121} and refs. therein).}
\label{fig:1}
\end{figure}

\section{The model and method}
We consider a chain of two-level (spin) units with nearest
neighbor XY couplings. The Hamiltonian is given by
\begin{equation}
H=-\sum_{i=1}^{N-1}[J(X_{i}X_{i+1}+Y_{i}Y_{i+1})+h(i)Z_{i}] \label{eq1},
\end{equation}
where $h(i)$ is an external field applied along the $z$ direction. $J$ is the coupling between nearest neighboring sites and is set
to $J=1$ for simplicity.
$X_{i},Y_{i}, Z_{i}$ denote the Pauli operators acting on spin $i$, and $N$ is the
total number of sites.
We consider a natural configuration of the spin chain
with open ends. As the total Hamiltonian preserves the excitation number,
the evolution of the initial state will remain within the initial
excitation subspace. Here we consider the system to be in the
\textquotedblleft one-magnon\textquotedblright\ state, where the total
number of up spins is one.
The \textquotedblleft one-magnon\textquotedblright\ state contains a single excitation in the
system, in accordance with the low-excitation condition. The model is the hard-core
boson limit of the Bose-Hubbard Hamiltonian for \textit{spinless} bosons, which has been implemented
experimentally in optical lattices \cite{QC,bloch}.
In this limit, we may have only one boson per site, so that one can encode an effective $1/2$ spin variable as the presence/absence of a boson at each site \cite{lewenstein}. This system can be prepared by loading a one-dimensional Mott insulator with one boson per site \cite{fukuhara2013}, and the using single-site addressing techniques to remove single bosons from the chain, by using laser \cite{Weitenberg} or electronic beams \cite{wuertz2009}. In principle, by calibrating the intensity and time duration of the beam, it should be possible to create arbitrary superpositions of $|0\rangle$ and $|1\rangle$ states. 
Specifically, we assume to start with a pure state of the form
$|0,0,...,0\rangle$,  where the occupied is $|0\rangle$ and
the empty is  $|1\rangle$ at each site. And then we apply the external beam to, let's say, the
first site. Assuming that the interaction with the beam occurs
on a timescale much shorter that the characteristic timescale
associated to the system Hamiltonian (and this is a very reasonable
assumption), the state of the system would be a pure state of the form
$(a|0\rangle+b|1\rangle)|0,...,0\rangle$, and tracing makes no difference.
Note that all above discussions are based on the hard-core boson limit, where
each site is only allowed to have either 1 or 0 boson because of strong
on-site repulsion \cite{bloch}. Then,
 a parabolic external field of the form
\begin{equation}
h(i) = 4h_{m}[(i-i_{F})^{2}/(i_{L}-i_{F})^{2}-(i-i_{F})/(i_{L}-i_{F})]
\label{eq:external-h}
\end{equation}
can be realized by an additional harmonic confinement \cite{dalfovo} (or even provided by the same laser beam that produces the lattice \cite{bloch}), where $h_{m}$
is the intensity of the external field. Fig. \ref{fig:1} shows the sketch of an experimental setup. The field distributes symmetrically with respect to the chain centre, and
is set to be zero at the end points $i_{F}=1$ and $i_{L}=N$ (this condition is not strictly necessary because energy is defined modulo a global constant).
This open end configuration can be realized as discussed in \cite{Wang20121}. The feasibility of the single-site-resolved addressing and control of individual spin states in an optical lattice has been experimentally demonstrated recently \cite{Weitenberg}. This allows the preparation of the system in an arbitrary configuration and the subsequent readout, making these system as promising prototypes for quantum memories.


We can diagonalize the Hamiltonian $H$ in the one magnon subspace so that $H_{d}=W^{\dag }HW$. The evolution
operators are therefore expressed by
\begin{equation}
U(t)=W\exp [-itH_{d}]W^{\dag },
\end{equation}%
where the time-independent $W$ is a unitary transformation between $H$ and its diagonal form $H_d$, and is obtained numerically.

Initially we prepare a state such that the $j$th spin, as our target site, is in the state $\left\vert \phi (0)\right\rangle
=\left\vert 1\right\rangle$, whereas all other spins are in the state $\left\vert 0\right\rangle $. The whole
spin chain will be in a product state $\Phi (0)=\left\vert 00\cdots 0\right\rangle \otimes \left\vert 1\rangle \otimes
|00\cdots 0\right\rangle $. The fidelity at time $t$, measuring the survival probability of the initial
state $\left\vert \phi (0)\right\rangle $, can be defined as $F=\sqrt{\left\langle \phi (0)\right\vert \rho (t)\left\vert \phi
(0)\right\rangle } $%
, where $\rho (t)$ is the reduced density matrix of the state at site $j$.


\section{A breathing pattern and dynamical quantum memory}
Let us start by
considering the first spin as our target site.
\begin{figure}[t]
\centering
\includegraphics[width=0.7\columnwidth]{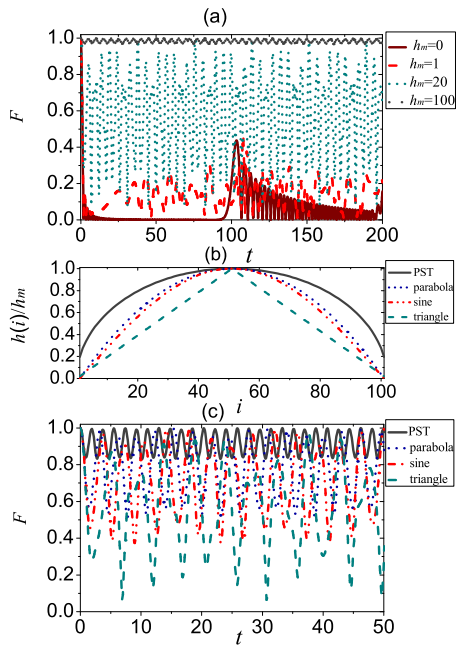}
\caption{(Color on line) (a) The fidelity as a function of time $t$ for different
$h_{m}$, where $N=100$. (b) External field functions $h(i)$. (c) Corresponding fidelities
as a function of time $t$ for different $h(i)$, $h_{m}=20$. }
\label{fig:2}
\end{figure}
Fig.~\ref{fig:2}(a) plots the fidelity versus time \textbf{($t\in[0, 200]$)}
for different field intensities $h_{m}$, for a chain of length $N=100$.
The figure shows that, in the absence of the external field ($h_{m}=0$),
the fidelity $F(t)$ drops from 1 to 0 in a very short time, as expected for a {\em primitive}  $XY$ model when $h_{m}$ in eq. (\ref{eq:external-h}) is zero.
Although there are oscillations in time, the fidelity never comes
back to the initial value again. In the presence of weak external field,
the fidelity remains small. When $h_{m}$ exceeds a certain value (e.g., $h_{m}=10$), the fidelity
revives at discrete time intervals $t_{0}$'s. In the case that $h_{m}$ exceeds 100, $F(t)\approx1$ and oscillates from 1 to 0.97 extremely rapidly.
This is an interesting quantum breathing pattern, where the
breathing frequency increases with the intensities of the external field.
This behaviour can be explained by noticing that the Hamiltonian (\ref{eq1}) is equivalent to a fermionic lattice hamiltonian \cite{AD}
in the presence of an external parabolic potential. In this case, single particle orbits in phase space belong to two different classes, as discussed in \cite{Pezze2004}:  \textit{closed orbits}, corresponding to oscillations around the parabolic potential minimum; and \textit{open orbits}, for particles performing Bloch-like oscillations at the sides of the parabolic potential, where the local slope is steep enough. In our case, when the target site is at one of the two ends of the chain, $i=i_{F}/i_{L}$, a semi-classical approximation shows that the latter behavior occurs roughly when $h_{m}>4$.
The target site or the first spin can therefore
serve as a quantum memory to store the spin state $\left\vert 1\right\rangle$, in the sense that $\left\vert 1\right\rangle$
is stored in time and retrieved at these discrete time moments $t_0$'s.

The parabolic external field is an idealization. Experimental implementation may vary from
this configuration, which might alter the breathing pattern. We thus examine other symmetric
configurations in Fig. \ref{fig:2}(a) for the external fields. These fields are in a
PST configuration:  $h(i)/h_{m}=2\sqrt{i(N+1-i)/(N+1)}$; the parabola as defined in
 Eq. \ref{eq:external-h}; sine: $h(i)/h_{m}=\sin [(i-1)\pi /(N-1)]$; and triangle
 $h(i)/h_{m}=2/(N+1)\min (i,N+1-i)$.  These functions are renormalized such
 that they have the same maximal values. Fig. \ref{fig:2}(b) shows that these configurations result in similar memory effects, and interpolation between two configurations gives similar results.
 This suggests that all these configurations,
smaller on the ends and bigger in the middle, and their interpolations will generate the dynamical
quantum memory. In other words, {\it the breathing pattern and corresponding
quantum memory is fault-tolerant against variations of the configurations of the external fields.}

\begin{figure}[t]
\centering
\includegraphics[width=0.6\columnwidth]{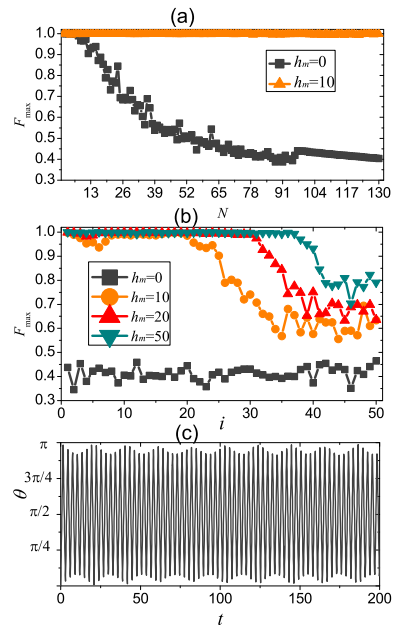}
\caption{(Color on line) (a) Maximum fidelity $F_{max}$ as a function of
the chain length $N$, with and without the external
field ($h_{m}=0,10$), respectively. The time window is [$100,1000$]. (b) Maximum fidelity $F_{max}$ as a function of
the storage location $i$, for various amplitudes of the external
field. (c) The values of the angle $\theta$ that correspond to the maximal fidelities $F_{max}$.}
\label{fig:3}
\end{figure}

Let us now address the dependence of the fidelity on the chain length.
In Fig.~\ref{fig:3}(a), we show the maximum fidelity $F_{max}$
in the time window $t\in[100, 1000]$ for
various chain lengths from $N=2$ to $N=130$, with and without the external field.
(The time interval [0, 100] is excluded since long-lived memory is more interesting).
In the absence of the external field, the fidelity $F_{max}$ decreases rapidly with $N$.
On the other hand, the presence of the parabolic field ensures good fidelities.
As shown in this figure, when $h_{m}=10$,
$F_{max}\approx 1$ and is independent of the chain length. Therefore, without loss of generality we will use $N=100$ in the following demonstrations.

The reason for the $N$-dependence is interesting. The condition $h_{m}>4$ is independent of the total length $N$ of the chain, because the actual amplitude of the parabolic potential 
depends on $N=i_{L}-i_{F}+1$, such that
when we compare chains of different lengths at fixed $h_{m}$, we also change the
potential amplitude (in a way that nothing else changes).

We now consider what happens if the state $\left\vert 1\right\rangle $ is stored at an
arbitrary site $i$, which is not the first site of the chain.
In Fig.~\ref{fig:3}(b) we show the maximum fidelity $F_{max}$ as a function of storage site $i$,
for different field intensities (for symmetry reasons it's enough to consider $i=1, 2, ...50$ for $N=100$).
As expected, $F_{max}(<0.5)$ is small in
the absence of external field.
$F_{max}$ is evidently enhanced for all sites in the
presence of the field, and is good when $i$ is small.  It also remains stable and then drops suddenly at a certain site $i_0$.
When $h_{m}=10$, the {\em{dropping}} site is $i=21$.
It implies that a quantum memory should be made in the sites near by one of ends of the chain.
Bigger field intensities, as shown before, also act to
broaden the storage region. For instance, the dropping
becomes $i_{0}=38$ when $h_{m}=50$ such that any of the 38 sites can be used as a quantum memory.
The semi-classical approximation shows that $i_0$ roughly satisfies $(\sqrt{h_{m}/4})(N+1-2i_{0})>(N-1)$, which is consistent
with the numerical observations in Fig.~\ref{fig:3}(b).

In general, a quantum memory requires one to store or memorize an arbitrary
quantum state $\left\vert \phi (0)\right\rangle =\alpha \left\vert
0\right\rangle +\beta \left\vert 1\right\rangle $ at the $j$th site. This can be done by including the magnon-zero
state $\left\vert 00..0\right\rangle$, which is the ground state of the whole system.
The corresponding fidelity is $f(t)=\left\vert \left\vert \alpha
\right\vert ^{2}+e^{i\theta(t)}\left\vert \beta \right\vert ^{2}F(t) \right\vert$, where $F$ is the fidelity for the state $\left\vert
1\right\rangle $. The fidelity $f$ of an arbitrary state $\left\vert \phi
(0)\right\rangle $ is given by $F$ and the angles $\theta$ in Fig.~\ref{fig:3}(c).  The stored arbitrary state can be retrieved when $\theta \approx 0$ such that
$f(t)= \left\vert \alpha\right\vert ^{2}+\left\vert \beta \right\vert ^{2}F(t) $, whose minimum is $F$. The relation $f\ge F$ always holds for an arbitrary state.

\section{Quantum memory against defects and noises} 

\begin{figure}[t]
\centering
\includegraphics[width=0.85\columnwidth]{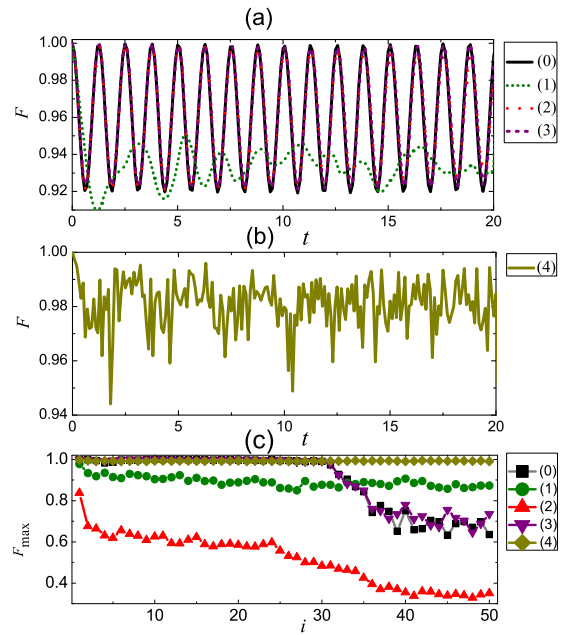}
\caption{(Color on line) (a) (b) The average fidelity as a function of time $t$ for
different defects and noises. (c) Maximum average fidelity $F_{max}$ as a function of
the storage site $i$,  for the five cases discussed in the text. The time window is [$100,1000$], where $h_{m}=60$. }
\label{fig:4}
\end{figure}

Let us now consider influences of
general fabrication defects and time-dependent random noises on the
fidelity of the quantum memory. We will analyze four cases of defects and noises.
(0) The primitive Hamiltonian (\ref{eq1});
(1) Band broadening, which comes from an additional term $H_{\text{site}}=\epsilon \sum_{i}$rand$(i)Z_{i}$ in the primitive Hamiltonian. Here rand$(i)$ is the random function in the interval $[-1,1]$;
(2) Random coupling due to fabrication errors, where the coupling in Hamiltonian (\ref{eq1}) is replaced by $J+\gamma \text{rand}(i)$, with random error $\gamma \text{rand}(i)$ ;
(3) Next-nearest neighbor contribution:  $J_{i,i+2}=\mu J$.
(4) Randomness in the coupling, where $J$ is replaced by a time-dependent random coupling $J+\eta\text{rand}_{\tau}(i)$;
The time-dependent random term simulates a noisy environment. The parameters $\epsilon
,\gamma ,\mu $ and $\eta $ are the strength of these perturbations.
Cases (1) and (2) are static perturbations due to fabrication defects.
Case (3) occurs when considering pseudospins based on charge degrees of freedom or
considering the dipole-dipole interaction \cite{Ronke}. Case (4) considers
that the random function rand$(i)$ is fixed in short time interval $\tau$
and it is randomly different for each time interval. We then calculate (thousand times) average density matrices and the corresponding fidelity.  In
Fig.~\ref{fig:4} (a)(b), we plot the time evolution of the fidelity for the five
cases, where the state $\left\vert 1\right\rangle $ is at the first site
and $h_{m}=60.0$. The parameters $\epsilon ,\gamma ,\eta
$ and $\mu $ are set to be $10\%$ of normal values of $h_m$ and $J$: $\epsilon =6.0,\gamma
=\mu =\eta =0.1$. Fig.~\ref{fig:4}(a) shows that the fidelity does not
change significantly with considered perturbations. Fig.~\ref{fig:4}%
(b) corresponds to the time-dependent noises $(\tau=0.1)$, and shows that
the maximum oscillates between 0.94 and 1.

In Fig.~\ref{fig:4} (c) we plot $F_{max}$ of different storage sites for
the above defects or noises. It is noticeable that $F_{max}$
does not change significantly in our time window except case 2 with random coupling. This implies that dynamical memory  can only be held at the first site in presence of random coupling.  
In case 3, the defects cause slight deviations from the original. The band broadening decreases $F_{max}$ sightly more. But
time-dependent noises surprisingly enhance $F_{max}$'s for most sites.
For most perturbations, as long as they are below the ten percent thresholds, many sites can be a very good candidate for quantum memory, and fault-tolerant against the
perturbations.
A possible explanation for the fault-tolerance may be the topological stability of the quantum XY model \cite{DeGottardi}. The interaction can be
mapped into the one-dimensional $p$-wave superconductor, whose states in the topologically non-trivial phase regime are robust against local fluctuations. 

\section{Experimental parameters}
The dimensionless parameters used in this paper can be easily converted into a dimensional form, 
for a direct comparison with those of current experiments with ultracold atoms. First of all, our choice $J=1$ corresponds to a dimensional time variable $(\hbar/J)t$. 
In the experiments, the one-dimensional chain considered here can be realized by means of an optical lattice in the tight-binding regime, where $J\simeq1.43 s^{0.98}e^{-2.07\sqrt{s}}E_{R}$ \cite{Gerbier} with $E_{R}=\hbar^{2}k_{L}^{2}/(2m)$ being the recoil energy, $k_{L}$ the wave vector of the lasers creating the lattice, and $s$ a dimensionless parameter. For typical experimental parameters (see \textit{e.g.} \cite{Weitenberg}, with $s=23$), $\hbar/J\simeq5\times10^{-2}$ so that the characteristic timescale of breathing pattern  discussed here is of the order of few tens of milliseconds, a typical timescale for ultracold atom experiments. In addition, our parabolic potential corresponds to a harmonic trapping of frequency
$\omega = \sqrt{8 J h_{m}/m}/(d(N-1))$. Then, for $h_{m}=1$, $N=100$, and the above lattice parameters, 
yield $\omega\approx 2\pi\times 1$ Hz, is again in the range of typical experimental values.

\section{Conclusions}

We have described a breathing pattern for the XY model trapped in a parabolic potential and have given a physical explanation for the interesting phenomena.
Based on this pattern, we have proposed a scheme to realize high-fidelity
dynamical quantum memory in a 1-D optical lattice. The maximum
fidelity increases with the field intensities and depends on the sites where the memory is located. High-fidelity quantum
memory can be realized in the large field regimes and at sites nearby an end of the chain. The maximum fidelity is independent of the chain
length. The scheme is also fault-tolerant against shapes of the external fields, various defects and noises, including those from manufacturing defects or bath. The setup is obviously more multifunctional than the one qubit memory. 
For instance, we could turn off the parabolic external field as in refs. \cite{DeGottardi,Gerbier} such that the stored state in the first qubit can be automatically transferred to other locations under spin chain dynamics
transfer along the $XY$ chain. The setup is a natural combination of quantum memory and quantum state transfer \cite{Bose, Liu}.

\begin{acknowledgements}
This material is based upon work supported by NSFC(Grant No. 11005099), Fundamental Research Funds for the Central
Universities (No. 201313012),
he Basque Government (grant IT472-10), the Spanish MICINN (Project  No. FIS2012-36673-C03-03) and the Basque Country University UFI (Project No. 11/55-01-2013), the NSF PHY-0925174,
DOD/AF/AFOSR No.~FA9550-12-1-0001.

\end{acknowledgements}

\end{document}